\documentclass[]{article}
\usepackage[english]{babel}
\usepackage{graphicx,enumerate}
\usepackage[tbtags]{amsmath}

\begin{document}
\title{Information and Interpretation of Quantum Mechanics}
\author{Marcin Ostrowski\footnote{e-mail: marcin.ostrowski@p.lodz.pl or mostrowski111@gmail.com}}
\maketitle

\begin{abstract}
This work is a discussion on the concept of information. We define here information as an abstraction that is able to be copied.
We consider the connection between the process of copying information in quantum systems and the emergence of the so-called classical realism.
The problem of interpretation of quantum mechanics in this context is discussed as well.
\end{abstract}

\section{Introduction}

What is information? In the literature, there is a multiplicity of definitions and approaches, such as statistical-syntactic, semantic and
pragmatic. The first of these examines statistical properties of messages, the second one attempts to explore the meaning of the content
contained in messages, while the third one determines their value in decision-making process.

\noindent
Milestone for the development of classical information theory was the work of C. Shannon from the 1940s' \cite{lit3}. Due to its practical importance in technology (e.g. telecommunications), Shannon's theory is widely presented in literature and is described extensively in many textbooks
(e.g. \cite{lit2, lit2a, lit2b}).

\noindent
Let us recall the main features of Shannon's information, which we are going to rely on in further sections:

\begin{itemize}
\item {Shannon entropy $H$ is a probability measure.  It describes the statistical characteristics of the processes such as telecommunication signal without going into the meaning contained in the transmission. It is a measure of uncertainty which is generated during the implementation of the process.  In other words, it is the a priori uncertainty as to the form that the implementation takes before we get to know it.}

\item {In Shannon's theory, there is a distinction between entropy $H$ (selfinformation) and information $I$ (transinformation).
This distinction is made by introducing the idea of the telecommunications channel. Message (process $X$) generated by the sender distorted by noise (process $Z$) gets to the recipient (process $Y$). Transinformtion $I(X,Y)$ is part of ignorance about the process $X$ which is removed when we know the process $Y$.}
\end{itemize}

\noindent
Interest in the notion  of information is particularly vivid  among physicists. For example, it is worth to mention the decades of discussion on the relationship between information and thermodynamic entropy \cite{lit_termo, lit_term2}.

\noindent
In recent times, range of applicability of the concept of information was extended to quantum systems. In quantum information theory there is a  distinction between classical information (as available locally ability to perform work) and quantum information (as nonlocal correlation which can be used to perform tasks such as teleportation)\cite{lit_horodecki, lit_preskill}. Despite all the modifications basic features of quantum information and Shannon's information are still the same. Both are  probabilistic measures of correlation between processes (or states of physical systems).

\noindent
The relationship between the concept of information and the problem of interpretation of Quantum Mechanics is also studied in the literature \cite{lit_horod2}. Despite significant progress in recent years, some questions still seem to be open. Some words have not been spoken directly. What really is the information? What is the relationship between the concept of information and the concept of the state of a physical system? Does the concept of information can be useful  for us in the discussions on the problem of interpretation of quantum mechanics? This work is another attempt to find answers to these questions.

\section{Information as an abstraction}

But what exactly is information? What are its main properties? If we were to answer that question in  ``our own words'' we would probably say that it is an abstract content contained in the transmission (of a message) or (looking more broadly) in a physical system. It is something
``immaterial'' included in the matter.  It is best to discuss this with an example. Consider the information that is normally stored in computer's memory (it can be a data file or a program). If the computer stores it in RAM, then it is represented by electric charges stored in microscopic
capacities (presence of the charge means the logic state ``1'', absence of the charge means logic state ``0''). Does this mean that the information can be identified with the electric charge or its distribution?  The answer is ``no''.  We can move the data from RAM to another medium. For example, we can save them to a magnetic disc (where information is stored in the distribution of magnetic domains), send them using a network wire (where information is encoded as a modulated electromagnetic signal), or print them on a punched card. As a matter of fact, these data need not be represented by a  binary system (i.e., by a string of ones and zeros). For example, if they represent a text, then on a paper printout we will have graphic marks representing the letters instead of zero-one sequence of bits. What we mean by the concept of information is not a specific
physical realization (e.g. record in the RAM) but rather a collection of many different realizations on different types of physical media, and encoded in different ways. Abstractness of information means that we can not identify it specifically with any of them.  Information must therefore
be transferred between different media.

\noindent
This situation reminds us of the definition of a rational number known from school. A colloquial knowledge of these numbers simply says that they are fractions, such as 3/4. They are pairs of integers. One integer (i.e. 3) represents the counter, and the other (i.e. 4) - the denominator.
If we constructed the definition of a rational number in this way, the fractions 3/4, 6/8 or 75/100 would represent different numbers. So, the definition of a rational number states that it is a class of abstraction. Within this class, all forms are equal and represent the same abstract number. In the case of information, the situation is the same. We can move it from one medium to another, and change its form of encoding.
However, the abstract content that is transferred  is still the same.

\noindent
Is the ability of information to be transferred from one medium to another all we need? How can we be sure that after a few transfers between various media the content of a message remained unchanged?  The best method of checking this is by comparing the content of a new medium with
the one stored in the old.  If they are both the same, then we can be sure that there were no errors.  Then, both copies can be considered as identical to each other and representing the same abstract information. This means that in addition to transfer operation, information must be
subject to the operation of comparison.\footnote{For rational numbers we have bringing to a common denominator procedure. It allows to assess whether the two fractions ($a/b$ and $c/d$) represent the same value.}

\noindent
However, the operation of  comparison makes sense if, in some circumstances it gives a positive, while in other circumstances, a negative result.
If there is only one copy of the message (without any possibility of duplication) comparing it with another message yield always a negative result. This means that instead of a transfer operation (during which the content is removed from the old medium and appears on the new one)
information must undergo a copy operation\footnote{
If after comparison of two unrelated messages, it appears that they are accidentally identical then, in fact, this is equivalent to copying them - there are two identical messages instead of one. Thus, we see that comparability is basically a concept close to  ``copyability''.} 
(during which the content appears in the new medium, while retaining the old).\footnote{
In analogy to computer terminology where file-copy operation (``copy'' - ``paste'' options) means that we get two copies of the file but after the
file-moving operation (``cut'' - ``paste'' options) we still have only one copy of the file in a new location.}
We can therefore say that:
\begin{quote}
{\it
Information is an abstraction contained in the physical system which is able to being copied \footnote{
In the literature available to the author, the definition similar to this was not found. Discussion on ``what is information?'' can be found in many works (for example: \cite{lit_podobne})}
}
\end{quote}

\noindent
``Copyability''\footnote{Copyability - ability to be copied} of information seems to be such an important feature that it should not be regarded as one of the many (additional) properties but as an essential feature. Everything that in the broad sense is defined as information has the
ability to reproduce (copy). In many cases, the number of copies may be enormous, making that information become a kind of ``public good''. Copying takes place not only when a user explicitly copies a file in the computer from one drive to another using the ``copy'' button. It often occurs when we are not aware of it. Our whole life is based on the content which is  subject to multiple copying. This applies to knowledge of historical facts, the insignificant events of our lives, and the prices of goods in stores, to name but a few.

\noindent
``Copyability'' makes information able to be shared. If we know a fact we can keep it in our memory and simultaneously share it with others. If the information was subject only to the transfer, instead of copying, then it would be impossible. In this case, either we could only remember the fact (without being able to share it with others), or we could pass it to someone else (although it could be just one person!) but then we would have to forget about it ourselves. Thanks to the ``copyability'' information about a fact is propagated among more and more people.
\footnote{Sometimes we also have to deal with objects that are ``transferable'' but not ``copyable''. For example, the quantum state. Its transfer from one object
to another (sometimes called teleportation) is possible while the copying is impossible (nocloning theorem).}
\footnote{
It should be stressed that when we talk about copying of information we mean replication of abstract contents without duplication of the form
(medium). The medium can be different each time, as described previously.}

\subsection{Subjective Probability}

Therefore, one can ask a question: What is the relationship between what is written above (about information as an abstraction  undergoing copying)
and information treated as a probabilistic measure? In order to resolve this issue, a closer look at the concept of probability
is needed.

\noindent
What is probability? What it really means (how to understand the statement) that the probability of a certain event is, say, 1/2? Generally, there are two approaches. The first of them (objective approach) gives a clear interpretation which states that if a random experience is repeated many
times then in 50\% of cases we should obtain the desired result. But there is also the subjective approach, according to which the probability may not mean the objective frequency of occurrence of results but may specify a subjective opinion (knowledge) on a topic depending on the currently available data \cite{pstwo1, pstwo2}. Let us illustrate this with a simple example:

\vspace{0.1cm}
\noindent
{\bf Example.}\,\,{\sf
Consider the transmission of a text file in English between computers. The text is displayed on the screen of the recipient's computer and then
read by the recipient. Suppose that this file contains the text of a well-known literary work. If it has been read, say, $n$ the first signs of
this work, what is the probability that as the $n+1$ sign there appears, for example, letter ``a''?

\noindent
Imagine two different recipients of that work. Let us assume that the first recipient has read the work many times before and knows it so well that he/she is able to recite it from memory. From the point of view of such a recipient, the  appearance of the letter ``a'' is a certain event (if at $n+1$ position there actually is letter ``a'') or an impossible event (with any other letter in the position of $n+1$). For such a recipient, this process is therefore completely deterministic in nature, and the entropy of each next symbol will be equal to zero.

\noindent
Imagine a different receiver (number two) which, unlike the first one, did not know the content of this work but knew that the text was in English.
Then, he/she assigns the probability of the letter ``a'' according to the statistics of the English language. From his/her point of view, it is a stochastic process.\footnote{For example, such a receiver can be a device carrying out the compression of text before sending it to the network, optimized for data compression in the English language.}}
\vspace{0.1cm}

\noindent
The approach presented here sometimes arouses controversy. However, many scientists
accept a subjective approach. We also stand for this interpretation. We assume that the probability depends on the observer and it is a measure of the observer's assumptions based
on previous experience.

\begin{quote}
{\it
In our work we assume that the probability of an event does not have an objective character but may be associated with a priori knowledge of
which an observer (receiver) has  about its reasons or effects.}\footnote{
It does not mean that we postulate the existence of observers as a different class of entities (other than physical systems subject to observation). For us, the observer is just another physical system. We argue that the concept of information makes sense only if we consider it as a correlation between the two sub-systems. Which one is called "observer" and which one is called "observed system" in fact is not important.}
\end{quote}

\noindent
Note that the assessment of probabilities in the example made by the recipient number 1 and number 2 stem from the fact that both (directly
or indirectly) previously encountered with the realization of the same process. For the first recipient, it was exactly the same work. For the
second recipient, they were different texts written in the same language.\footnote{In this case the ``process'' means a specific language.} In both
cases we deal with copying of realization. Very often we encounter such situations in real world. Multiple copies of the same realization reach the
observer at different moments of time.  Earlier copies (not always faithful) become a knowledge base which is used to assess the probability of copies
coming later. So we find here the relationship between the (subjective) probability and information as an abstraction that undergoes copying.

\begin{quote}
{\it
The assigning of the realization of the process specified value of subjective  probability is associated with an earlier copying of information
about this realization.}
\footnote{
Perhaps it would be better if we used here the concept of conditional probability rather than the concept of subjective probability. The main postulate of this section is the following: in the nature only conditional probabilities have sens. In practice, the probability of an event $A$ (labeled as $P(A)$) is always conditional probability. This means that for each event $A$, we always have $P(A)=P(A/X)$ where $X$ are some conditions which occurrence can be detected only by observation (that is, in fact, by copying information). For example for coin toss ($A$=HEAD, $B$=TAILS, $P(A/X)=P(B/X)=1/2$) $X$ means: that coin was fair (e.g. it has no property of badminton shuttlecock failing always one side) and that toss was fair (e.g. coin has not been dropped freely from a 5cm height, in a horizontal position). To conclude that conditions $X$ took place an observation (and de facto copying of information) is required. Who wants to play in a casino where the roulette wheel is invisible to the players?}
\end{quote}

\noindent
In Shannon's theory, information is defined on the basis of the concept of probability. Hence, it can be regarded as a concept secondary in relation to probability. In this work, in some sense, we propose to reverse the order. We consider information as a primary concept, defined as an abstract content which undergoes copying. The probability of an event is a secondary concept associated with information (with one of its copies) which an observer has about this event.

\noindent 
In Shannon's theory, information (transinformation) is the relation between two processes or subsystems. The copying process is therefore a process of establishing this relation.

\section{Copyability in Quantum Theory}
\subsection{Quantum Copier}

In Shannon's theory, we speak about information in the context of its transport through the telecommunication channel. The process $X$ is defined as the sent  message, and the process $Y$ , as a received message. However, keeping all the mathematical properties of these processes,
we can look at the issue differently.  If, instead the telecommunication channel, we consider a copying machine (copier) then the process $X$ can be seen as the original, and in the process $Y$, as the obtained copy.  $I(X,Y)$ is then a measure of fidelity of copying process.

\noindent
In this section, the copying of information is looked upon in the context of quantum mechanics.
As is well known, there is no possibility of copying arbitrary quantum states (no-cloning theorem). However, it is possible to faithfully copy the information encoded in orthogonal states of the quantum system. Let us look at an operation of information copying between two quantum subsystems now. This operation will be called a quantum copier. 

\noindent
Let us consider a physical (quantum) system that stores one bit of information. Let logic state ``0'' correspond to the quantum state $\vert0\rangle$ and logic state ``1'' correspond to the quantum state $\vert1\rangle$. States $\vert0\rangle$ and $\vert1\rangle$ are orthogonal to each other, normalized and form the basis of a quantum system. Assignment of logical values for selected physical states can be identified with the establishment of a ``communication protocol''.

\noindent
Consider two such systems which will be called, $A$ and $B$, respectively. Let us construct a quantum operation that copies the states ``0'' and
``1'' from the system $A$ to system $B$ according to the following scheme:
\begin{eqnarray}
\label{eq_kop1}
\vert0\rangle_A\vert\textrm{pm}\rangle_B\rightarrow\vert0\rangle_A\vert0\rangle_B\\
\label{eq_kop2}
\vert1\rangle_A\vert\textrm{pm}\rangle_B\rightarrow\vert1\rangle_A\vert1\rangle_B\\
\label{eq_kop3}
\vert0\rangle_A\vert\textrm{um}\rangle_B\rightarrow\vert0\rangle_A\vert1\rangle_B\\
\label{eq_kop4}
\vert1\rangle_A\vert\textrm{um}\rangle_B\rightarrow\vert1\rangle_A\vert0\rangle_B
\end{eqnarray}

\noindent
Equations (\ref{eq_kop1})-(\ref{eq_kop4}) fully define the operation of the copier as a unitary operation acting over $H_A\otimes H_B$. States $\vert\textrm{pm}\rangle_B$ and $\vert\textrm{um}\rangle_B$ are orthogonal each other. Results of copying are correct if system $B$ (medium) is initially in state $\vert\textrm{pm}\rangle_B$ (pure medium). Otherwise, when system $B$ is in state $\vert\textrm{um}\rangle_B$ (unprepared medium), an error occur. Such way of copier's functioning is dictated by the requirement of unitarity of (\ref{eq_kop1})-(\ref{eq_kop4}) operation.\footnote{Analysis of the copier work in the context of energetic properties of the process of information copying was carried out in \cite{moj1}.}\,\footnote{
The scheme presented above can be generalized to copying more states:
\begin{equation*}
\vert i\rangle_A\otimes\vert\textrm{pm}\rangle_B\rightarrow\vert i'\rangle_A\otimes\vert i\rangle_B
\end{equation*}
where $i, i'=0, 1, \dots n$. Note that in our generalized scheme, the state of the system $A$ is changing during the copying from
$\vert i\rangle_A$ to $\vert i'\rangle_A$. This type of change is admissible if states  $\vert i'\rangle_A$ are mutually orthogonal.
The essence of information copying is to be able to determine what the initial state of $A$ was, knowing the final state of $A$ or final state
of $B$. What quantum states we use to encode logical states is completely arbitrary, provided that the potential recipient knows the convention
that we use. As previously mentioned, it is a question of establishing a kind of communication protocol intelligible to the interested. At this
point, is also worth noting that using the same letter $i$ for mark states-symbols in $A$ (i.e. $\vert i\rangle_A$) and states-copies in $B$
(i.e. $\vert i\rangle_B$) does not mean that these states must have the same physical nature (for example, the projection of spins on the same
axis). This is just to emphasize the fact that they encode the same logical values. Generally, those states ``live'' in different subspaces and
their physical nature may be entirely different. What is important is that both sets of states are mutually orthogonal, i.e. the copy operation
must be a mapping of the orthogonal states of $A$ to the orthogonal states of $B$.}

\subsection{Measurement as a copying of information}

\subsubsection{Von Neumann measurement}

Let us recall briefly the idea of von Neumann measurement \cite{lit_pom}. For this purpose, let us consider a quantum system initially being
in a state  $\vert\Psi\rangle$. This state can be presented as superposition of eigenstates $\vert s\rangle$ of some observable $\hat{A}$
in the form: $\vert\Psi\rangle=\sum_s c_s\vert s\rangle$. The Von Neumann model introduces the concept of a measuring apparatus (which is also
a quantum system) being initially in the state $\vert A_0\rangle$. The effect of measurement is the transition of the state of the measured
system and the measuring system, according to the scheme:
\begin{equation}
\label{vonNeumann}
\vert A_0\rangle\otimes\vert\Psi\rangle
\rightarrow
\sum_s c_s\vert A_s\rangle\otimes\vert s\rangle
\end{equation}
We can see an analogy between the operation of the copier (\ref{eq_kop1})-(\ref{eq_kop4}) with the von Neumann measurement (\ref{vonNeumann}). If we associate system $A$ (storing the original message) with the measured system
and system $B$ with the measuring apparatus, then states $\vert s\rangle$ correspond to the states $\vert i\rangle_A$, states of the apparatus
$\vert A_s \rangle$ correspond to the states $\vert i\rangle_B$, and initial state of apparatus $\vert A_0\rangle$ can be identified with the
state $\vert\textrm{pm}\rangle_B$.

\noindent
Indeed, during the measurement it comes to the process of copying information from object $A$ to the measurement system (observer) $B$. After measuring, the information about the state of $A$ is contained both in the system $A$ and in the observer $B$. Thus, there is the same information in two subsystems and this is exactly the sense of information copying. If sombody (system $C$) wants to obtain knowledge about which of the states $\vert i\rangle$ has obtained $B$, then he/she have two options: ``asking $B$'' about their result or perform independently of $B$ the next measurement on $A$ (the same observable), as a result of which they he/she certainly get the same result as $B$.

\subsubsection{The problem of uniqueness of the measurement}

It turns out that decomposition of the non-separable state obtained after measuring (right side of Eq.~(\ref{vonNeumann})) is not unambiguous
\cite{lit_pom2, lit_pom3}. Other decompositions are possible (using other bases): 
\begin{equation}
\sum_s c_s\vert A_s\rangle\otimes\vert s\rangle=\sum_r d_r\vert A_r\rangle\otimes\vert r\rangle
\end{equation}
This has been interpreted as an ambiguity of the von Neumann measurement. It is not known whether the measurement (projection) was made on the
states $\vert s\rangle$ or $\vert r\rangle$. Does this mean that the operation of our copier is also ambiguous? The answer is ``no''. If useful information was recorded in the states $\vert0\rangle_A$, $\vert1\rangle_A$ (in the second notation there are states $\vert s\rangle$) then after copying it is stored in the system $B$ in the states $\vert0\rangle_B$, $\vert1\rangle_B$ (in the second
notation $\vert A_s\rangle$). Measurement using a base $\vert A_r\rangle$ corresponds to the attempt to read information stored in the $B$ by using states ($\vert\tilde0\rangle_B$, $\vert\tilde1\rangle_B$) other than the correct ``states-symbols'' ($\vert0\rangle_B$, $\vert1\rangle_B$). It is easy to check that this would lead to a loss of information copied earlier to $B$. The condition for the correct operation of the copier is that the initial state of $A$ is one of the states $\vert i\rangle_A$ and further reading of the information copied to $B$ is made using the states $\vert i\rangle_B$. 
It is a question whether, during copying, the predetermined ``communication protocol'' is respected. We will return to this issue in the next sections.

\section{Information and interpretation of Quantum Mechanics}
\subsection{Copying to multiple subsystems}

In this section we examine a situation where the initial state of $A$ is a superposition of states-symbols in the form:
\begin{equation}
\label{eq_kot}
\vert\Psi\rangle_A=\alpha\vert0\rangle_A+\beta\vert1\rangle_A
\end{equation}
We will successive copy information from the system $A$ (according to the scheme described by equations (\ref{eq_kop1})-(\ref{eq_kop4})) to many systems
$B_1$, $B_2$, \dots, $B_n$ initially being in the states $\vert\textrm{pm}\rangle_{B_i}$. For each copying process the states $\vert0\rangle_A$
and $\vert1\rangle_A$ will be the states-symbols. States-copies will be denoted by $\vert0\rangle_{B_i}$, $\vert1\rangle_{B_i}$.

\noindent
After making the first copy, from system $A$ to $B_1$, the state of system $A\,B_1$ takes the form:
\begin{equation}
\label{eq_jeden_bis}
\vert\Psi_{AB_1}\rangle=\alpha\vert0\rangle\vert0\rangle+\beta\vert1\rangle\vert1\rangle.
\end{equation}

\noindent
If, in the same way, other observers ($B_i$) perform the copy (from $A$ or any of the previous $B_i$) the final state of system will be:
\begin{equation}
\label{eq_wielu}
\vert\Psi_{AB_1\dots B_n}\rangle=\alpha\vert0\rangle\vert0\rangle\dots\vert0\rangle+\beta\vert1\rangle\vert1\rangle\dots\vert1\rangle
\end{equation}
This type of copying of information to many observers, using exactly the same states-symbols will be called multi-copying.\footnote{
States of this type are called von Neumann chains \cite{lit_pom}. However, this concept is not used in the context of copying of information.}

\noindent
State (\ref{eq_wielu}) has an interesting property, namely, if we make the measurement (reduction of state) on $A$ or any of the $B_i$ in base $\vert0\rangle$, $\vert1\rangle$\footnote{This is of course a simplified notation denoting $\vert0\rangle_A$, $\vert1\rangle_A$ in the case of
system $A$ and $\vert0\rangle_{B_i}$, $\vert1\rangle_{B_i}$ in the case of the one of the systems $B_i$} then with probability $\vert\alpha\vert^2$ we obtain state of the whole system in the form $\vert0\rangle\vert0\rangle\dots\vert0\rangle$ or with probability $\vert\beta\vert^2$ we obtain state of the whole system in the form $\vert1\rangle\vert1\rangle\dots\vert1\rangle$. Moreover, if some systems $B_i$ are measured by another observers and each of them performs measurement using base $\vert0\rangle$, $\vert1\rangle$ then either everyone obtain $\vert0\rangle$ or all obtain $\vert1\rangle$. Despite the initial superposition of the $A$ everyone receive the same version of reality.

\begin{quote}
{\it
When all subsystems performing copying of information between themselves use the same set of states-symbols, then they always share the same version of reality (even if the initial state is a superposition of states-symbols).
}
\end{quote}

\subsection{When multi-copying occurs in nature - examples}

An example of the multi-copying of information can be a well-known interference experiment with electron and two slits. Let $\vert 0\rangle$ mean that the electron passed through the left slit when $\vert 1\rangle$ means that it passed through the right slit.  We choose the initial state of the electron in the form of Eq.~(\ref{eq_kot}) which means that the slit through which the electron moved to the other side of the shield is not specified. However, what we are interested in is which slit the electron actually passed through. To test it, we put a light source near slits in such a way that observing the scattered photons it is possible to determine near which slit the scattering has occurred. Let us denote the photon scattering states: $\vert 0_f\rangle$ - photon scattered in the left slit, $\vert 1_f\rangle$ - photon scattered in the right slit. 

\noindent
Let us suppose that the beam of light which illuminates the area near the slits is so strong that very many photons can be scattered with the electron when it is close to one of the slits. Then each of the large number $n$ of scattered photons is carrying conclusive information through which slit the electron transition occurs. The state of the system electron plus $n$ photons can be written as:
\begin{equation}
\label{eq_vizja}
\vert\Psi_{e,f_1,\dots,f_N}\rangle=\alpha\vert 0\rangle\vert 0_{f_1}\rangle\dots\vert 0_{f_N}\rangle+
\beta\vert 1\rangle\vert 1_{f_1}\rangle\dots\vert 1_{f_N}\rangle,
\end{equation}
which is an analogy with the Eq.~(\ref{eq_wielu}).

\noindent
In this experiment, information about the way chosen by an electron (system $A$) is multicopied to photon states (systems $B_i$). All copies have
been made using the same states-symbols. The result is that all the photons share the knowledge about the slit which was chosen by the electron.
Since the initial state of the electron was a superposition of states-symbols, the whole system finally reaches a state of (\ref{eq_vizja}) describing
the mutual entanglement of all particles (subsystems).

\subsection{When does it come to reduction of state?}

Let us continue discusion about the experiment with two slits. Suppose that in the experiment we have a photon detector recording the scattered photons and able to observe through which slit the electron passed.
This detector changes its quantum state as follows: it reaches the state $\vert 0_d\rangle$ if it registers a photon scattered at the left slit or it reaches state $\vert 1_d\rangle$ if the scattered photon is observed at the right slit. Then the state of a system consisting of an electron, $n$ photons and the detector can be presented in the form:
\begin{equation}
\label{najsplot}
\vert\Psi_{e,f_1,\dots,f_N,d}\rangle=\alpha\vert 0\rangle\vert 0_{f_1}\rangle\dots\vert 0_{f_N}\rangle\vert 0_d\rangle+
\beta\vert 1\rangle\vert 1_{f_1}\rangle\dots\vert 1_{f_N}\rangle\vert 1_d\rangle
\end{equation}

\noindent
This type of analysis can be carried further by introducing other quantum subsystems (for example a person  recording the behavior of the detector, etc.). Each of the subsystems can be assigned two quantum states describing two alternative courses of experience and the whole system
can be assigned state of type (\ref{najsplot}).
So where is the reduction of state? If we continue our discussion further, will we always be doomed to a superposition of two alternative versions of reality?

\noindent
This question involves the widely discussed for decades problem of interpretation of quantum mechanics. During  this time many competing concepts  were created, trying to deal with this problem. As a reminder  let us mention the most common: Copenhagen, many worlds \cite{lit_pom3, lit_manyworlds}, Bohmian \cite{lit_Bohm}, Consistent Histories \cite{lit_hist1, lit_hist2, lit_hist3}, modyfied dynamics (as for example GRW model \cite{lit_GRW}). Each of them has  group of supporters and opponents (discussion of this issue can be found in review articles \cite{lit_przegladowe}).

\subsection{Multi-copying: arbitrariness in setting  the border}

As is well known, many physicists have skeptical attitude to the Copenhagen interpretation, resulting in many competing interpretations. In our work we also reject it.
We also assume that
\begin{quote}
\begin{center}
{\it
all systems in nature are quantum systems.
}
\end{center}
\end{quote}
It may seem quite a natural assumption, since all physical objects are composed of microscopic particles governed by the laws of quantum mechanics.
This means that not only the microscopic systems, but also their macroscopic observers are quantum systems. But then we meet the basic and widely
discussed problem: when does it really come to reduction of state? In the remainder of the work, we will focus on a discussion of this issue.

\noindent
Recall that in the second section we opted for the subjective interpretation of probability. If so, the state of a quantum system, usually interpreted statistically (probability amplitude), should also have a subjective character. It could be a measure of knowledge about the system possessed by a
specific observer. In our work, we accept this interpretation:
\begin{quote}
{\it
We assume that the state of a quantum system is a measure of subjective knowledge of the observer about the system.\footnote{epistemological
status vs. ontological status of quantum state}
}
\end{quote}

\noindent
Different observers can assign the same system to different states (one of them pure, the other mixed) depending on the extent of knowledge they possess about the system. In that case, the process of reduction of state is subjective too.  One observer may conclude that reduction occurred, while from another observer's point of view no reduction took place.  We assume that 
\begin{quote}
{\it
reduction of the wave function of the object takes place from the perspective of an observer who is gaining new information about the object.
}
\end{quote}
The system evolves unitarily if it does not exchange information with environment or with the observer. We could say that unitary evolution describes
the change of ``representation'' of the same knowledge of the system over time. If this knowledge is altered as a result of inflow of new data to the
observer then brand new ``representation'' is needed, which we describe by the process of reduction of state.

\noindent
Let us return to the analysis presented in Section 4.1. In our understanding state described by
Eq.~(\ref{eq_jeden_bis}) should be interpreted as follows:
\begin{equation}
\label{eq_interpret}
\vert\Psi_{AB_1}\rangle=\sum_i\alpha_i\vert\textrm{\scriptsize $A$ is in state i}\rangle
\vert\textrm{\scriptsize $B_1$ knows, that $A$ is in state i}\rangle.
\end{equation}

\noindent
From the point of view of  $B_1$, measurement was carried out, i.e. the state $A$ has undergone irreversible reduction to the resulting state
$\vert i\rangle$. From the perspective of another (external) observer (which for convenience will be called system $C$), the situation is completely different. There has been no reduction. State (\ref{eq_interpret}) of system $AB_1$ still corresponds to the superposition of all possible scenarios (results). This is due to the fact that from the point of view of system $C$, the transition from state (\ref{eq_kot}) to state (\ref{eq_jeden_bis})
does not involve the acquisition of any new information about the system $AB_1$ (there is no interaction between $C$ and $AB_1$). This transition is solely the result of knowledge possessed by $C$ about the initial state (\ref{eq_kot}) and the known a priori evolution of the system $AB_1$ (unitary
copy operation).

\noindent
Where the reduction of state takes place, depends on where we put the borderline between what belongs to the observer and what belongs to the observed system. In the reasoning leading to Eq.~(\ref{najsplot}) all systems (i.e. electron, photons and detector) were considered as part of the observed system. However, if we consider the detector as an integral part of the observer, then we can assume that, after it is reached by any of the photons, reduction of state occurs, and thus, description of the situation by means of Eq.~(\ref{najsplot}) simply has not a raison d'etre. We can do this if we are sure that the communication between us and the detector is reliable, which means that the observation made by the detector is equivalent to the observation made by ourselves. However, if the communication between the detector and us is distorted (e.g. as a result of interactions with the environment) then errors occur and our assumption is not correct.

\noindent
Analogously, the first photon scattered by electrons can be included in the measurement system. Here, the reduction of state occurs when the photon and the electron interact with each other and, thus, state (\ref{eq_vizja}) has no raison d'etre. In textbook descriptions of this experiment, it is often assumed that the interaction between the electron and the photon is equivalent to the (approximate) measurement of the position of the electron that destroys the original superposition. It is assumed tacitly that the photon will certainly fall to the detector, and that its state will still include information about where the scattering takes place (i.e. there will be no scattering with other particles) and that the detector will perform the proper measurement of photon observable (which will not destroy the encoded information on the location of the electron).

\begin{quote}
{\it
It does not matter where the border between the observer and the observed system is placed {\rm if} information obtained by the observer
(as a result of the reduction occurring on this border) is in every situation exactly the same.
}
\end{quote}

\noindent
It does not matter where we put the borderline, as long as we obtain descriptions of reality that are not in contradiction to each other.\footnote{
The problem: ``where  to put the border between the observer and the physical system''  was discussed since the dawn of Quantum Mechanics. The thesis that the border should be moveable was presented in \cite{lit_granica}.}
However, it is necessary that all systems use during copying the same set of states-symbols (multi-copying). This is possible only if, for quantum systems found in nature, such a distinguished set of states-symbols really exists.

\noindent
The existence of the distinguished quantum states have been considered in the works of other authors \cite{lit_zurek, lit_zurek2, lit_zurek3}, where the notion of {\it preferred pointer states} was introduced i.e. quantum states very willingly monitored by the environment. This permits unambiguous measurement ({\it preferred observable}) and leads to the {\it environment-induced superselection rules}. Our view is similar but not identical. We discuss this matter in the next
section.

\section{Physical space and copying of information}

Consider a quantum system composed of a number of subsystems. Treating quantum mechanics as a formal mathematical theory, the Hamiltonian describing the interaction between these subsystems can be chosen from a variety of operators.\footnote{Of course, such an operator should meet a number of formal requirements such as: self-adjointness, limiting the spectrum from the bottom, etc.}
In general, the interactions described by such an operator do not lead to the processes of copying information. Moreover, even if copying of information occurs, states-symbols may be different each time.

\noindent
We may ask how often it comes in nature to the situation (described in previous) which for each sub-system there is a universal set of states-symbols
used during copying of information to other subsystems? Do the Hamiltonians that describe the systems occurring in nature favor any specific set of state-symbols? Is there any ``communication protocol'' preferred by the nature?

\noindent
There is also another question. For processes of copying information occurring in nature, how do we know which states are states-symbols?
In practice, nature has established a ``communication protocol'', so from the observer point of view, it may be unknown a priori. Then, how does the observer know which states carry information?

\noindent
Looking at the example mentioned in previous section, note that states-symbols are always associated with a fixed location of subsystems (particles) in physical space. It seems reasonable to argue that: 

\begin{quote}
{\it
For all subsystems (particles) occurring in nature, there is a universal set of states-symbols. The elements of this set are spatially localized
states.
}
\end{quote}

\noindent
When it comes to information copying between the particles then states-symbols are related to their specific locations. So we can informally identify a physical space with a universal set of states-symbols common to all physical particles. This does not mean that we identify states-symbols with
the eigenstates of position operator $\hat{x}$, i.e. with the the Dirac delta states. We do not treat the position as a distinguished observable.
Interactions between particles are never point type interactions (exchange of intermediate bosons). Moreover, when we use photons for localization of objects, we are doomed to resolution of the wavelength of light.\footnote{In the case of massive particles, it is resolution of de Broglie
wavelength.} Therefore, speaking of the spatially localized states, we mean states whose position is determined with an accuracy of the wavelength used for the location of the particles. In experiments carried out at the macro level (e.g., in ``everyday life'' scale) using visible light
($\lambda<1\mu m$) such a statement is correct. Of course in other cases, where the wavelength $\lambda$ is comparable to the size of the experimental area, this term is not correct. However, to simplify the language, such a states we can call briefly the ``spatially localized states''.

\subsection{Multi-copying and decoherence}

Processes of information copying are ubiquitous in nature. A copy of the information is often placed in a system which is a part of the external environment $E$. It is easy to check that, if the system $A$ (carrying the original information) initially was in the state (\ref{eq_kot}) then after
copying  $A\rightarrow E$,  its state (from the perspective of the experimenter $C$) becomes a mixed state\footnote{Pure state can be identified with the situation when the observer knows the actual copy of the information about the system. Mixed state of the system means that probably unknown for observer copies of the information about the system exist in the external environment. It means that system $A$ is better correlated with $E$ than with observer $C$.} in the form:
\begin{equation}
\label{eq_a_miesz}
\rho_{A}=\vert\alpha\vert^2\vert0\rangle_{A\,A}\langle0\vert+\vert\beta\vert^2\vert1\rangle_{A\,A}\langle1\vert.
\end{equation}

\noindent
Let us return to the quantum copier and the problem of unambiguity of its work (section 3.2) Let us focus for a while on the source of information which has been denoted as a system $A$. Useful information (undergoing copying) was encoded in a binary form in the states $\vert0\rangle_A$ and $\vert1\rangle_A$. As we have said, this can be identified with the establishment of a ``communication protocol'' which requires encoding of information only in these two states, rejecting other as meaningless. This raises several questions. First, what is the degree of arbitrariness of such a ``protocol'' (the choice of states-symbols)? Second, what would it mean if a state of the system $A$ arose as a superposition of states-symbols? In order to answer these questions, imagine for example that the system $A$ is a sealed container which holds a coin
and an apparatus (or person) for coin-tossing. If the coin comes up heads, we have ``0'', if the coin comes up tails, we have ``1''. Imagine also that in the wall of the container is a hole protected by a flap. When the flap is closed, the information never leaves the container (a unitary evolution
takes place). When we open the flap, we can look inside and observe if it the outcome is "0" or "1". Thus, this system in an automatic way respects the established ``protocol'', because there are only two possible results of the coin-tossing: either heads or tails.

\noindent
Let us now consider whether $A$ may be in the quantum state (\ref{eq_kot}), i.e. a superposition of possible outcomes, and what it means. If we assume an objective interpretation of the state, then it can only mean a ``protocol'' error (e.g. the device is working improperly). However, we have accepted a subjective interpretation. Then, the state (\ref{eq_kot}) may mean that the tossing went well but the information about its result did not escape outside. That means neither we nor anyone else (e.g. environment) knows the result. The state (\ref{eq_kot}) is therefore the result of a purely theoretical prediction: we know the state before the tossing and we know how its (unitary) evolution was proceed. 

\noindent
The ``protocol'' can be interpreted as a principle, which states that ``when we throw a coin, it can come up either 0 or 1''. But what do we really understand by the word ``come up''? So it is better if the ``protocol'' states that ``when we look inside after the tossing, we should see the coin in
a state 0 or 1, never in any other''. In such situation state (\ref{eq_kot}) is not inconsistent with the ``protocol'', since it means that we have not looked inside yet.

\noindent
It is now time to put the fundamental question: what distinguishes states $\vert0\rangle_A$ and $\vert1\rangle_A$ from other states (their superpositions) i.e. what determines the form of the ``protocol''? Sometimes it is suggested that it is the environment that selects those states by
monitoring the measurement apparatus or directly the measured system (with the coin). However, it seems that this is not the cause. States $\vert0\rangle_A$ and $\vert1\rangle_A$ are distinguished by the fact that they are stable and they differ from each other by the {\bf location in space} of the object with large mass and size, which is the coin. In practice, both we (the observers) and environment, wanting to ``look into'' the container with a coin, may use particles like photons and electrons. As a result, we are forced to choose states-symbols in the form of states with a fixed spatial location. It does not matter who performs the measurement first (opens the flap, lets photons from the outside in and looks inside), we or the environment. Of course, in practice, the environment does it much faster, making the copies of information (``0'' or ``1'') propagate outwards in a large number of copies, which results in the transition of a pure state (\ref{eq_kot}) into a mixed state (\ref{eq_a_miesz}). But it is not the environment that chooses the base. If we do this first by using ``what nature has given'' we are also forced to choose base ``0'' and ``1'' as a natural ``protocol''.

\section{Local irreversibility}

The copying of information (such as operation (\ref{eq_kop1})-(\ref{eq_kop4})) is a reversible operation, i.e. there is an operation $U^{-1}$ reverse to it. This means that any process of information copying can be revoked.\footnote{
So the state $\vert0\rangle\vert0\rangle$ will go back into the state $\vert0\rangle\vert\textrm{pm}\rangle$ the state $\vert1\rangle\vert1\rangle$
will go into state $\vert1\rangle\vert\textrm{pm}\rangle$. We call it a revocation of copying instead of erasing, to not confuse it with erasure by the emission of heat \cite{landauer}}
So what guarantees that all the events described (such as the choice by the electron of a specific slit) actually took place if we can not exclude the possibility that, as a result of future evolution of the system, the operation $U^{-1}$ will take place, resulting in the revocation of all the copies? Even if information about the state of the system is copied, say 1000 times (up to 1000 subsystems $B_i$), it still does not become a part of physical reality. If (for example) a properly prepared experimenter who has access to all systems $B_i$ applies to them operation $U^{-1}$, then
all copies will be deleted. This type of experiments are described in the literature as the so-called ``quantum eraser''\cite{lit1, lit1b}.

\noindent
However, in practice, such cases do not happen (with the exception of experiments prepared in the laboratory). It is unlikely that all the particles
that hold copies of information re-enter into mutual interaction, which would result in the removing of all copies (operation $U^{-1}$). But as long
as these particles are ``close together'' in principle, this can not be ruled out.

\noindent
The case is different when photons take part in the experiment (process).\footnote{Since most processes in nature (including most of the processes important in biology) have an electromagnetic character, the participation of photons is widespread (even if they are only photons of infrared
radiation associated with heat emission).}.
Their fundamental property is the ability to move in a vacuum with a maximum speed attainable in nature, that is, the speed of light. So if a photon escapes into the open space then nothing will be able to catch it and force it to return to the experiment area.

\noindent
What does it mean from our point of view? If one of the systems $B_i$ to which the information is copied is a photon escaping in the free space, we will not be able to reverse the copying. Of course, this irreversibility is relevant locally, so this property should be called the ``local
irreversibility''. We can not be sure that we are not part of a larger experiment conducted on a cosmological scale crafted so that the photons finally come back to the area of the experiment and reversing of copying is realized. But it is a matter of the cosmological nature and it will not be considered here.

\noindent
However, one thing is certain. A property of ``local irreversibility'' guarantees that neither we nor anyone else in our nearest environment will
be able to reverse copying that has taken place. We can only to hope that no one else (in the cosmological scale) will do it either.

\noindent
Let us summarize our discussion:
\begin{quote}
{\it
If the state of a quantum system is copied to multiple subsystems, and some of these copies can not be (locally) removed, then it becomes an element of physical reality. This state is equated with the event.
}
\end{quote}

\noindent
The event is a quantum state which ``was copied'' so many times and in such forms that we can not reverse it. What counts here is not the number of copies, but the impossibility of removal them all. However, if we delete all copies of the information about a fact (for example, about
the way chosen by a particle), then this fact ceases to be an element of physical reality.

\noindent
What is most important in our approach is the statement that the event occurs when information about it is propagated in the environment. In our understanding, freely moving photon in the empty space is not considered as an event (or a set of events). In contrast, photon's emission from the atom, absorption by another atom or reflection from a mirror may be events. In such situations, states of both systems are changed and process of type  (\ref{eq_kop1})-(\ref{eq_kop4}) can take place. 
 
\noindent
The view presented here concerning the irreversibility, is perhaps a little naive. However, if we do not want to change the mathematical structure of quantum mechanics (introducing non-unitary mechanisms such as in the GRW model) then we are condemned to such explanations. Maybe a future quantum gravity will bring something more concrete to the case. Wider discussion of this issue, in the context of the existence of the arrow of time, can be found in \cite{lit_moj2}.

\section{Summary}

In the approach presented in this paper, we consider information as the most fundamental concept in nature. It is true that when we describe
the quantum systems, the concept of quantum state is used. But a quantum state is simply a mathematical way of expressing information we
have about the system. Information gained by an observer as a result of copying (either by direct measurement or by classical communication with another observer).

\noindent
The most important postulate in quantum mechanics is the principle of superposition. It says that if $\vert0\rangle$ and $\vert1\rangle$ are two states
that the system can take, then their superpositions $\vert\tilde0\rangle=\alpha\vert0\rangle+\beta\vert1\rangle$ and
$\vert\tilde1\rangle=\gamma\vert0\rangle+\delta\vert1\rangle$ are also correct physical states.
We do not deny in the slightest this postulate. However, it seems that due to the Hamiltonians found in nature, certain states are
distinguished as states-symbols in the process of information copying. 

\begin{quote}
{\it
According to the  postulates of quantum mechanics, all states are ``equal'', but because of the interactions occurring in nature, some of
them are ``more equal than others''
}
\end{quote}

\noindent
Distinguishing the states-symbols does not matter for a free particle. The case is different in the case of interacting particles. The interaction
identifies states-symbols with the places in space where it takes place with the accuracy of wavelength $\lambda$. For example, the work of organs
such as eyes is registration of photons, or, to be more specific, the measurement of directions from which the photons come. This measurement is most
often destructive (a photon is absorbed). So we can ask why we need the knowledge about the direction of a photon which no longer exists? Usefulness
of such a measurement follows from the fact that previously the information has been copied. The direction of the photon possesses encoded information
about the location (state-symbol) of the object that previously emitted or reflected this photon. 
It seems that distinguished role of ``physical space'' in the perception  of reality
  stems from the fact that localized states (``points of space'') correspond to states-symbols in the process of copying information. 

\noindent
In our work, we wanted to show that the world we live in, which in the daily observation appears to us as a classical, is in fact ruled at all levels
by the laws of quantum mechanics. Only a few additional features such as the existence of distinguished states-symbols and the ability of information
to create multiple copies leaking into the environment makes that world appears to us as classical. It seems that the classical world is described by
quantum mechanics plus several additional mechanisms related to the flow of information between systems.



\begin{thebibliography}{99}

\bibitem{lit3} C.~E.~Shannon, {\it A Mathematical Theory of Communication}, Bell System Tech. J., vol. 27 (1948);

\bibitem{lit2} F.~M.~Reza ``An Introduction To Information Theory'', McGraw-Hill, (1961);

\bibitem{lit2a} W.~W.~Mitiugow ``Physical foundations of information theory'', Moscow (1976);

\bibitem{lit2b} L.~Brillouin ``Science and Information Theory'', Academic Press Inc.,  New York-London (1962);

\bibitem{lit_termo} L.~Szilard, Z~f~Physik, {\bf 53}, 840, (1929);

\bibitem{lit_term2} C.~H.~Bennett, Int. J. Theor. Phys., {\bf 21}, 905, (1982);

\bibitem{lit_horodecki} J.~Oppenheim, K.~Horodecki, M.~Horodecki, P.~Horodecki and R.~Horodecki, {\it A new type of complementarity between quantum and classical information}, http://arXiv:quant.ph/0207025;

\bibitem{lit_preskill} J.~Preskill, http://www.theory.caltech.edu/~preskill/ph229;

\bibitem{lit_horod2} R.~Horodecki, M.~Horodecki and P.~Horodecki, {\it Quantum information isomorphizm: beyond the dillemma of Scylla of ontology and Charybdis of instrumentalism}, http://arXiv:quant-ph/0305024;

\bibitem{lit_podobne} M.~Mazur, ``Jakosciowa teoria informacji'', WNT, Warszawa 1970;

\bibitem{pstwo1} E.~Borel ``Probabilite et certitude'', Presses Universitaires de France, Paris (1961);

\bibitem{pstwo2} B.~V.~Gnedenko, ``The Theory of Probability'', MIR Publishers, Moscow (1982);

\bibitem{moj1} M.~Ostrowski, {\it Minimum energy required to copy one bit of information}, http://arxiv.org/abs/1004.4732;

\bibitem{lit_pom} 
 J.~Von~Neumann, ``Mathematische Grundlagen der Quanten Mechanik'', Springer-Verlag, Berlin (1932). English translation by
 R.~T.~Beyer. ``Mathematical Foundations of Quantum Mechanics'', Princeton University Press, Princeton (1955);

\bibitem{lit_pom2} W.~H.~Zurek, {\it Pointer Basis of Quantum Aparatus: Into What Mixture Does the Wave Packet Collapse?},
 Phys. Rev. D {\bf 24}, 1516, (1981);

\bibitem{lit_pom3} H.~Everet, {\it Relative State Formulation of Quantum Mechanics}, Rev. Mod. Phys. {\bf 29}, 454 (1957);

\bibitem{lit_manyworlds} J.~A.~ Wheeler, Rev. Mod. Phys. {\bf 29}, 463, (1973);

\bibitem{lit_Bohm} D.~Bohm, ``A suggested Interpretation of the Quantum Mechanics'', Cambridge University Press, (1987);

\bibitem{lit_hist1} R.~B.~Griffiths, {\it Consistent Histories and the Interpretation of Quantum Mechanics}, J.~Stat.~Phys. {\bf 36}, 219, (1984);

\bibitem{lit_hist2} R.~B.~Griffiths, ``Consistent Quantum Theory'', Cambridge University Press (2008);

\bibitem{lit_hist3} R.~Omnes, ``The Interpretation of Quantum Mechanics'', Princeton Univ. Press (1992);

\bibitem{lit_GRW} G.~C.~Ghirardi, A.~Rimini, T.~Weber, {\it Phys. Rev. D}, {\bf 34}, 470 (1986);

\bibitem{lit_przegladowe} M.~Tegmark, {\it The Interpretation of Quantum Mechanics: Many Worlds on Many Words?}, Fortschr. Phys. {\bf 46}, 855, (1998);
\\ J.~J.~Slawianowski, ``Przyczynowosc w mechanice kwantowej'', Wiedza Powszechna, Warszawa (1969);

\bibitem{lit_granica} N.~Bohr, Nature 121, 580-590 (1928);

\bibitem{lit_zurek} W.~H.~Zurek, {\it Decoherence and the Transition from Quantum to Classical-Revisited};

\bibitem{lit_zurek2} W.~H.~Zurek, {\it Decoherence, Einselection, and the Quantum Origins of the Classical}, http://eprints.lanl.gov.quant-ph/0105127;

\bibitem{lit_zurek3} W.~H.~Zurek, {\it Environment-induced superselection rules}, Phys. Rev. D {\bf 26}, 1862, (1982);

\bibitem{landauer} R.~Landauer, IBM J. res. Develop., {\bf 5}, 183, (1961),

\bibitem{lit1} P.~G.~Kwait, A.~M.~Steinberg, and R.~Y.~Chiao, {\it Observation of a ``Quantum Eraser'': a Revival of Coherence in a Two-Photon Interference Experiment}, Phys. Rev. A, {\bf 45}, 7729, (1992);

\bibitem{lit1b} X.~Y.~Zhou, L.~Mandel, {\it Induced Coherence and Indistinguishability in Optical Interference}, Phys. Rev. Lett., {\bf 67}, 318, (1991);

\bibitem{lit_moj2} M.~Ostrowski, {\it Information and the arrow of time},\\ http://arxiv.org/abs/1101.3070;

\end{thebibliography}
\end{document}